\documentclass{bmcart}

\usepackage{caption}

\usepackage[utf8]{inputenc} 

\def\includegraphics{}

\usepackage[numbers, sort&compress]{natbib}
\newcommand{\ignore}[1]{}
\usepackage{fancyhdr}
\usepackage[normalem]{ulem}
\usepackage[hyphens]{url}
\usepackage{hyperref}
\usepackage{graphicx}
\usepackage{multirow}
\usepackage{color}
\usepackage{float}
\usepackage{enumitem}
\usepackage{setspace}
\usepackage{amsmath}
\usepackage{arydshln}
\usepackage{booktabs}

\startlocaldefs
\endlocaldefs

\begin{document}

\begin{frontmatter}

\begin{fmbox}
\dochead{Methodology} 


\title{GRIM-filter: fast seed filtering in read mapping \\using emerging memory technologies}


\author[
   addressref={cmu1},
   email={jeremiekim123@gmail.com, calkan@cs.bilkent.edu.tr, onur.mutlu@inf.ethz.ch}
]{\inits{JSK}\fnm{Jeremie S.} \snm{Kim}}
\author[ 
   addressref={cmu1}, 
   email={dsenol@andrew.cmu.edu}
]{\inits{DS}\fnm{Damla} \snm{Senol}}
\author[
   addressref={cmu2}, 
]{\inits{HX}\fnm{Hongyi} \snm{Xin}}
\author[
   addressref={nvidia}, 
]{\inits{DL}\fnm{Donghyuk} \snm{Lee}}
\author[
   addressref={cmu1}, 
]{\inits{SG}\fnm{Saugata} \snm{Ghose}}
\author[
   addressref={bilkent},
]{\inits{MA}\fnm{Mohammed} \snm{Alser}}
\author[
   addressref={eth}, 
]{\inits{HH}\fnm{Hasan} \snm{Hassan}}
\author[
   addressref={tobb}, 
]{\inits{OE}\fnm{Oguz} \snm{Ergin}}
\author[
   addressref={bilkent}, 
]{\inits{CA}\fnm{Can} \snm{Alkan*}}
\author[
   addressref={eth, cmu1}, 
   email={omutlu@gmail.com}
]{\inits{OM}\fnm{Onur} \snm{Mutlu*}}


\address[id=cmu1]{
  \orgname{Department of Electrical and Computer Engineering, Carnegie Mellon University}, 
  \street{Forbes Avenue},                     %
  \city{Pittsburgh},                              
  \cny{USA}                                    
}
\address[id=cmu2]{%
  \orgname{Department of Computer Science, Carnegie Mellon University}, 
  \street{Forbes Avenue}, 
  \city{Pittsburgh}, 
  \cny{USA}
}
\address[id=nvidia]{%
  \orgname{NVIDIA Research},
  \city{Austin},
  \cny{USA}
}
\address[id=bilkent]{%
  \orgname{Department of Computer Engineering, Bilkent University, Bilkent, Ankara, Turkey}, 
  \city{Bilkent}, 
  \cny{TR} 
}
\address[id=tobb]{ 
  \orgname{Department of Computer Engineering, TOBB University of Economics and Technology}, 
  \city{Sogutozu}, 
  \cny{TR}
}
\address[id=eth]{ 
  \orgname{Department of Computer Science, Systems Group}, 
  \city{Zurich}, 
  \cny{CH}
}


\begin{artnotes}
\end{artnotes}

\end{fmbox}


\begin{abstractbox}

\begin{abstract} 
\parttitle{Motivation} 
Seed filtering is critical in DNA read mapping, a process where billions of
DNA fragments (reads) sampled from a donor are mapped onto a reference genome
to identify genomic variants of the donor.  Read mappers 1) quickly generate
possible mapping locations (i.e., seeds) for each read, 2) extract reference
sequences at each of the mapping locations, and then 3) check similarity
between each read and its associated reference sequences with a computationally
expensive dynamic programming algorithm (alignment) to determine the origin of
the read. Location filters come into play before alignment, discarding seed 
locations that alignment would have deemed a poor match. The ideal location
filter would discard all poor matching locations prior to alignment such that
there is no wasted computation on poor alignments.

\parttitle{Results} 
We propose a novel filtering algorithm, GRIM-Filter, optimized to exploit
emerging 3D-stacked memory systems that integrate computation within a stacked
logic layer, enabling processing-in-memory (PIM).  GRIM-Filter quickly filters
locations by {\em 1)} introducing a new representation of coarse-grained
segments of the reference genome and {\em 2)} using massively-parallel
in-memory operations to identify read presence within each coarse-grained
segment. Our evaluations show that for 5\% error acceptance rates, GRIM-Filter
eliminates 5.59x-6.41x more false negatives and exhibits end-to-end speedups of
1.81x-3.65x compared to mappers employing the best previous filtering
algorithm.
\end{abstract}




\end{abstractbox}
%

\end{frontmatter}



\section{Introduction} 


Our understanding of human genomes today is affected by modern technology's
ability to quickly and accurately determine an individual's entire genome. The
human genome is comprised of a sequence of approximately 3 billion bases that
are grouped into deoxyribonucleic acids (\emph{DNA}), but today's machines can
only identify DNA in short sequences (\emph{reads}). Therefore, determining a
genome requires 3 stages: 1) cutting the genome into many short fragments, 2)
identifying the DNA sequence of the fragment, and then 3) mapping the reads against
the reference genome in order to analyze the variations in the sequenced
genome. In this paper, we focus on improving stage 3, often referred to as
\emph{read mapping}.  Read mapping is performed computationally by \emph{read
mappers} after each read has been resolved into a known series of DNA. 

\begin{figure}[h]
    \centering
    \includegraphics[width=0.5\linewidth]{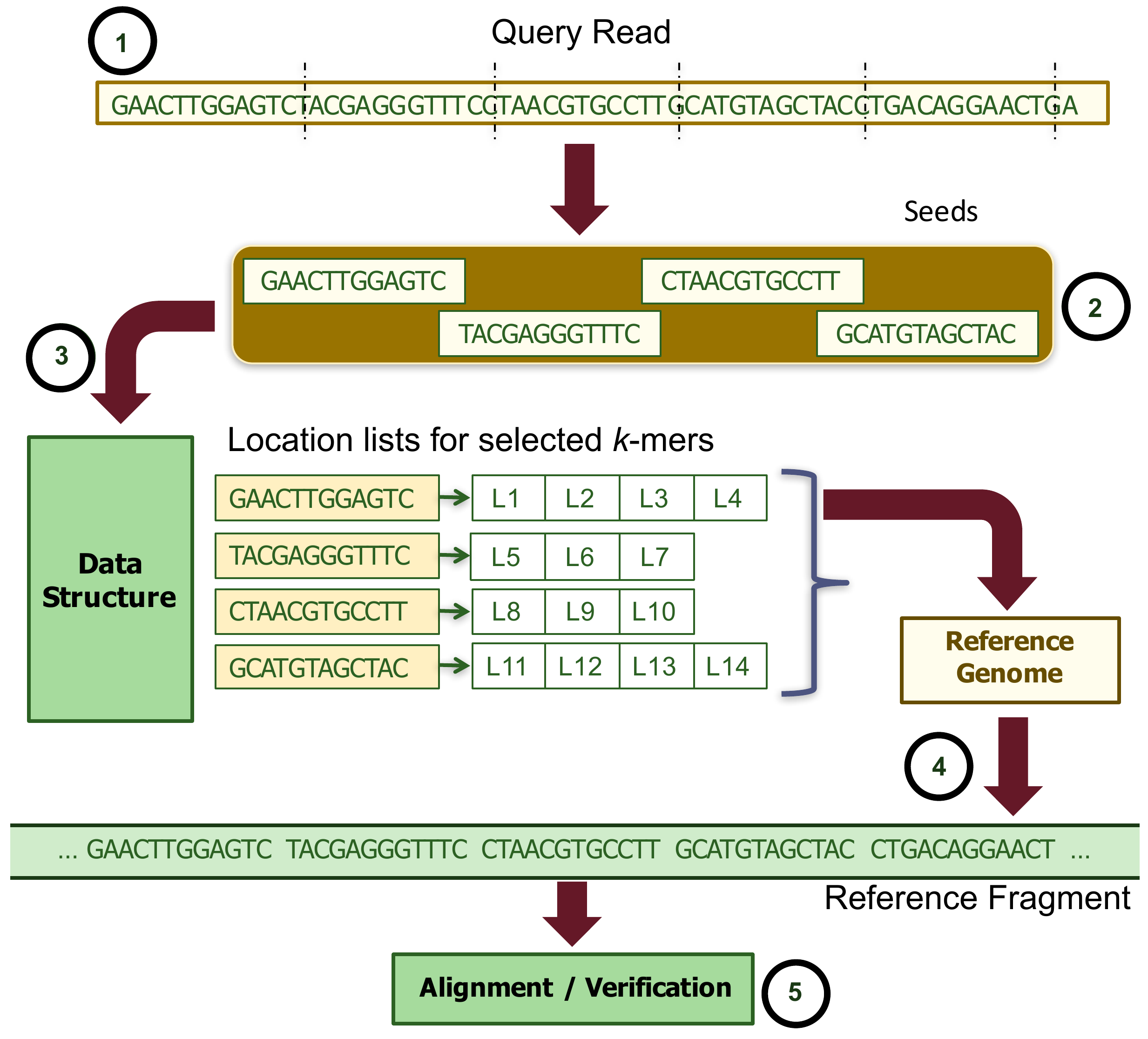}
    \caption{Flowchart of a seed-and-extend mapper} 
    \label{fig:fastHASH}
\end{figure}

We refer to Figure~\ref{fig:fastHASH} to briefly explain a class of read
mappers, seed-and-extend mappers. Seed-and-extend mappers attempt to find
locations in the reference genome that closely match each read sequence with
the following procedure. It 1) obtains a query read, 2) selects smaller
segments (i.e., seeds) of the read, 3) index a data structure with these seeds
to obtain a list of possible locations that would result in a match, 4) obtain
the sequence from the reference genome, and 5) align the read sequence to the
reference sequence with an expensive dynamic programming algorithm in order to
determine similarity. 

To improve performance on the runtimes of seed-and-extend mappers, we can
utilize {\em seed filters}, recently introduced by Xin et
al.~\cite{xin2013accelerating}. Seed filters efficiently determine whether
a candidate mapping location will result in an incorrect mapping \emph{before}
performing the computationally-expensive alignment step for that location.  As
long as the filter can eliminate possible locations faster than the time it
takes to execute alignment, the entire read mapping process will be
accelerated~\cite{xin2015shifted, xin2013accelerating}.  As a result, several
recent works have focused on optimizing the performance of seed
filters~\cite{tran2015amas, xin2015shifted, xin2013accelerating,
xin2015optimal, alser2016gatekeeper, alser2017magnet}.


The onset of seed filters has resulted in a shift of the performance
bottleneck to filtering but filters still require large amounts of memory
bandwidth to process and characterize each of the candidate locations. We
attempt to reduce the time spent in filtering and present a new algorithm,
GRIM-Filter, to efficiently filter locations with high parallelism. We observe
that the characteristics of GRIM-Filter reflect an algorithm well-suited for
implementation on 3D-stacked memory and evaluate GRIM-Filter on our in-house
3D-stacked memory simulator. 

3D-stacked DRAM~\cite{AMD-HBM, lee2015simultaneous, AMD-R9-Graphics,
o2014highlights, altera-HMC-UG, HMC-sources} is an available and emerging
technology that integrates logic and memory in a 3D stack of dies with a large
internal bandwidth. This enables the bulk transfer of data from memory to a
logic layer that can perform simple parallel operations on the data. 

Whereas conventional computing requires the movement of data on buses between
core and memory, processing-in-memory (PIM)-enabled devices such as 3D-stacked
memory enable simple arithmetic operations in nearby memory with high bandwidth.
With carefully designed algorithms mapped for PIM, applications can often
be improved immensely as the relatively small bus between core and memory no
longer impedes the progress of computation on the data. 

\textbf{Our goal} is to develop a seed filter that exploits the high memory
bandwidth and processing-in-memory capabilities of 3D-stacked DRAM to increase
the performance of hash table based read mappers without sacrificing their high
sensitivity or comprehensiveness. 

To our knowledge, this is the \textbf{first} seed filtering algorithm that
accelerates read mapping by overcoming the memory bottleneck with PIM using
3D-stacked memory technologies. GRIM-Filter can be used with any read mapper, 
however, in this work we demonstrate the effectiveness of GRIM-Filter with a 
hash-table based mapper. 

\textbf{Key Mechanism.} GRIM-Filter provides a quick method for determining
whether a read will \textbf{not} match at a given location, thus allowing the
read mapper to skip the expensive alignment process for that location.
GRIM-Filter works by counting the existence of small segments of a read in a
genome region. If the count falls under a threshold, GRIM-Filter discards the
locations in that region before alignment. The existence of all small segments
in a region are stored in a bit vector which can be easily predetermined for
each region of a reference genome and retrieved when a read results in a
potential location to a given region. We find that this regional approximation
technique not only enables a high performance boost via parallelism but also
improves filtering accuracy over the state-of-the-art. 

\textbf{Key Results.} We evaluate GRIM-Filter qualitatively and quantitatively
against the state-of-the-art seed filter
\emph{FastHASH}~\cite{xin2013accelerating}. Our results show that GRIM-Filter
yields a \emph{5.59x--6.41x} smaller false negative rate (i.e., proportion of
locations that pass the filter, but result in a poor match) than the best
previous filter, and runs end-to-end \emph{1.81x--3.65x} faster than
\emph{mrFAST} with \emph{FastHASH} for a set of real genomic reads, when we use
a 5\% error threshold. We also note that as we increase the error rate, the
performance of our filter over the state-of-the-art also increases, thus making
our filter more effective and relevant for future generation error-prone
sequencing technologies.

%
%
 
\section{Motivation and Aim} 

Mapping the reads against the reference genome enables the analysis of the
variations in the sequenced genome, and with a higher throughput in mapping
sequences, more large-scale analyses are possible. The ability to deeply
characterize and analyze genomes on a large scale could change medicine from
reactive to a preventative and further personalized practice. In order to
motivate our method of improving the performance of read mappers, we pinpoint
the performance bottlenecks of modern-day mappers on which to focus our
acceleration efforts.  We find that across our dataset, mrFAST with
FastHASH~\cite{xin2013accelerating} still spends 15\% of computation time
aligning locations that are found to be a match, and 59\% of the time aligning
locations that are later discarded (i.e., \emph{false locations}). Our goal is
to implement a filter that reduces the wasted computation time spent aligning
false locations by quickly determining if a location will not match the read
and forgo the alignment altogether. The ideal filter would exhibit no
additional overhead and correctly find all false locations and shows the
potential to improve average performance of mrFAST on the same machine by
\emph{3.2x}.  We note that this speedup is primarily earned by reducing the
number of false location alignments, whereas most prior works gain their
speedup by implementing parts or all of the read mapper in hardware. These
works are orthogonal solutions, and could be implemented together with location
filters for additional performance improvement.

\section{GRIM-Filter} 
\label{sec:grim_filter}

We now describe our proposal for a new seed filter, GRIM-Filter. At a high
level, GRIM-Filter utilizes meta-data on short segments of the genome in order
to quickly determine if a read will \textbf{not} result in a match at that
genome segment. 

\begin{figure}[h]
  \centering
  \includegraphics[width=0.5\linewidth]{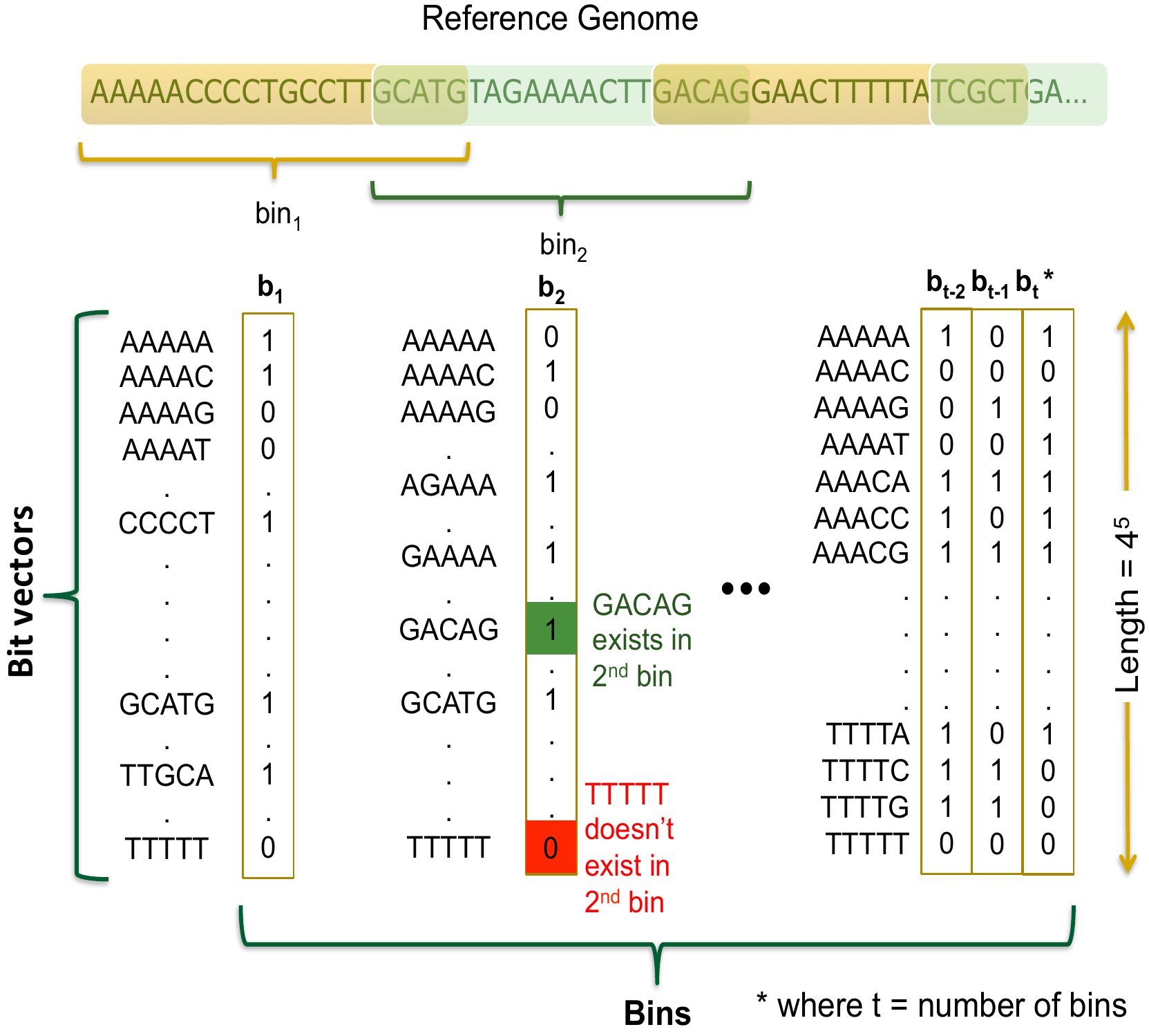}
  \caption{Data structure layout of bit vectors. Columns are indexed by bin-number-converted locations. Rows are indexed by the token hash value. In this figure, token size=$5$. }
  \label{fig:bitvectors}
\end{figure}

Figure~\ref{fig:bitvectors} shows a reference genome with its associated
meta-data. The reference genome is divided into short contiguous segments, on
the order of several hundreds of base pairs, which we refer to as \emph{bins}. GRIM-Filter
runs at the granularity of these bins, operating on the meta-data associated
with each bin.  This meta-data is stored in a \emph{bit vector} that stores
whether or not a \emph{token}, or small DNA sequence on the order of $5$ base
pairs, can be found within the associated bin. We refer to each bit as an
\emph{existence bit}. To account for all possible tokens of length $n$, each bit
vector must be $2^n$ bits in length, where each bit denotes the existence of a
particular token instance.  Figure~\ref{fig:bitvectors} highlights the bits of
two token instances of $bin_2$'s bit vector showing the existence of token
GACAG (green) with a $1$ and the lack of token TTTTT (red) with a $0$. 

Because these bit vectors are associated with the reference genome, the bit
vectors must only be generated once per reference and can be reused to map any
number of reads from other individuals of the same species. However, in order to
generate the bit vectors, the genome must be sequentially scanned for every
sequence of $n$ length tokens. If $bin_x$ contains the first base pair of a
token, the token's corresponding index of the associated $bit vector_x$ must be
set ($1$), but otherwise unset ($0$). These bit vectors can then be saved for
later reuse when mapping reads to the same reference genome used to generate
them. 

Before alignment, GRIM-Filter checks a read's potential mapping location by
operating on the bit vector of the bin holding the first base pair of that
read. This relies on the entire read being contained within a given bin,
requiring bins to overlap (i.e., some base pairs are contained in multiple
bins) as shown in Figure~\ref{fig:bitvectors}. 

GRIM-Filter uses these bit vectors in order to quickly determine if a match
within a given error rate is impossible. This is determined before running
alignment, the expensive dynamic programming algorithm in order to reduce the
number of unnecessary alignment operations. For each location, we 1) load the
bit vector of the bin containing the location, 2) operate on the bit vector (as
we will describe shortly) to quickly determine if there will be no match, and
3) discard the location if GRIM-Filter determines a poor match.  Otherwise, the
sequence at that location must be aligned with the read to determine the match
similarity. 

\begin{figure*}[t]
  \centering
  \includegraphics[width=0.9\linewidth]{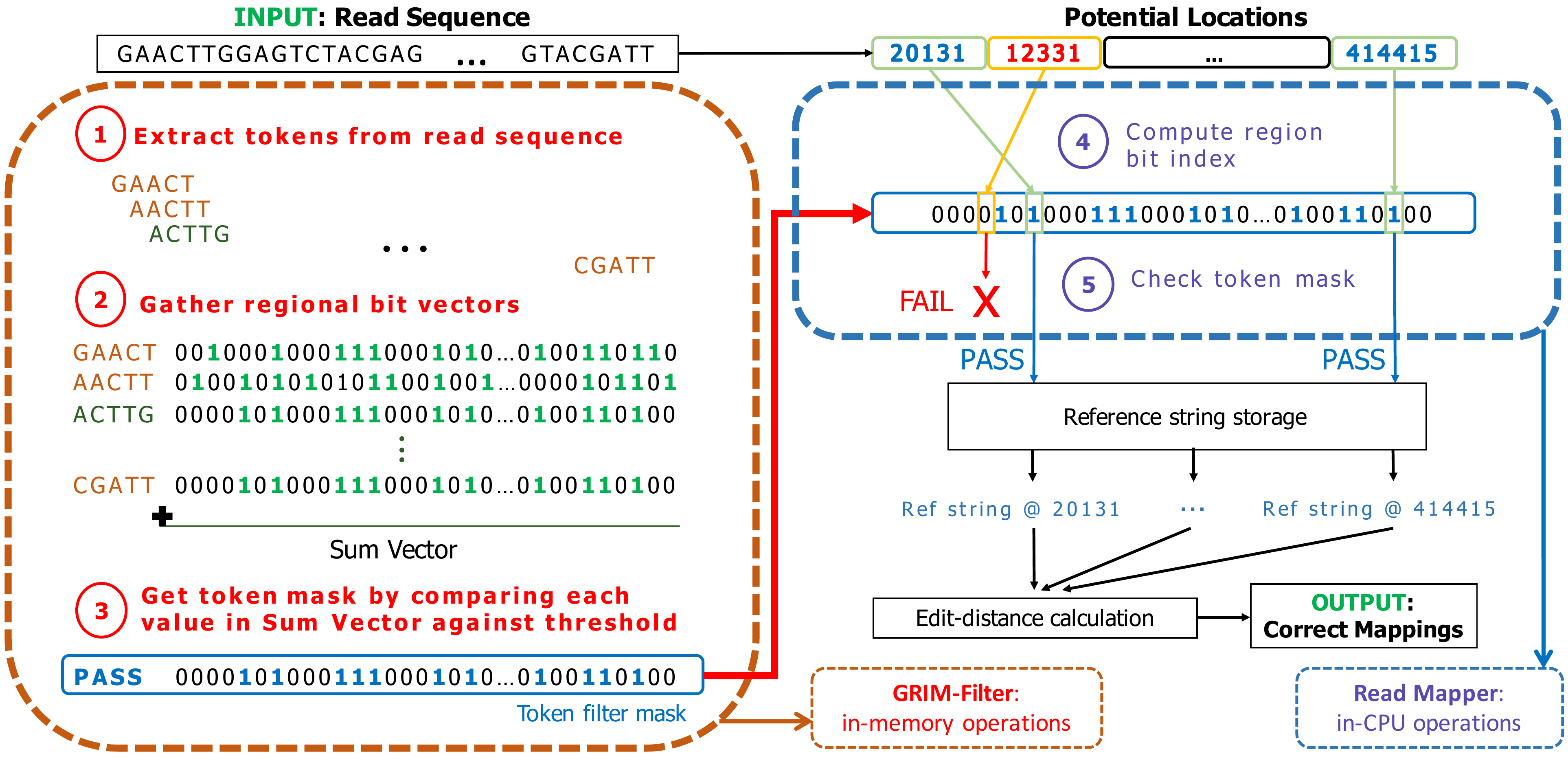}
  \caption{Flow diagram for our algorithm. GRIM-Filter takes in read sequences and generates filter masks on the 3D-stacked logic die using precomputed bit vectors, while the CPU filters locations, queried from the hash table, using the resulting filter mask. The locations of correct mappings and their edit distances are then returned to the user.} 
  \label{fig:fullAlgorithm}
\end{figure*}

Using the circled steps in Figure~\ref{fig:fullAlgorithm}, we explain in detail
how GRIM-Filter determines whether to discard a location $z$ for a read. 1)
GRIM-Filter extracts every token in the read and 2) accumulates their
respective existence bits from the bit vector. 3) The sums are compared to a
threshold (that we explain below), and set to 1 if it meets the
threshold, otherwise set to 0. When the read mapper is ready to align a read to
a a segment of the reference sequence, the read mapper must 4) determine which
bit to check against, and then 5) determine whether it should continue with
alignment or not. 

We now discuss in detail how to determine the threshold value to compare the
sum from step 3. A higher sum would represent a higher probability for the read
to match well within that bin, since a higher sum represents a higher number of
parts of the read being found in the bin.  However, intuitively, this may 
never confirm whether the read would match well or how well the match would be.
On the other hand, if the number of parts found falls below a certain
threshold, we can guarantee that the read will result in a poor match. 

\begin{figure}[t]
  \centering
  \includegraphics[width=0.4\linewidth]{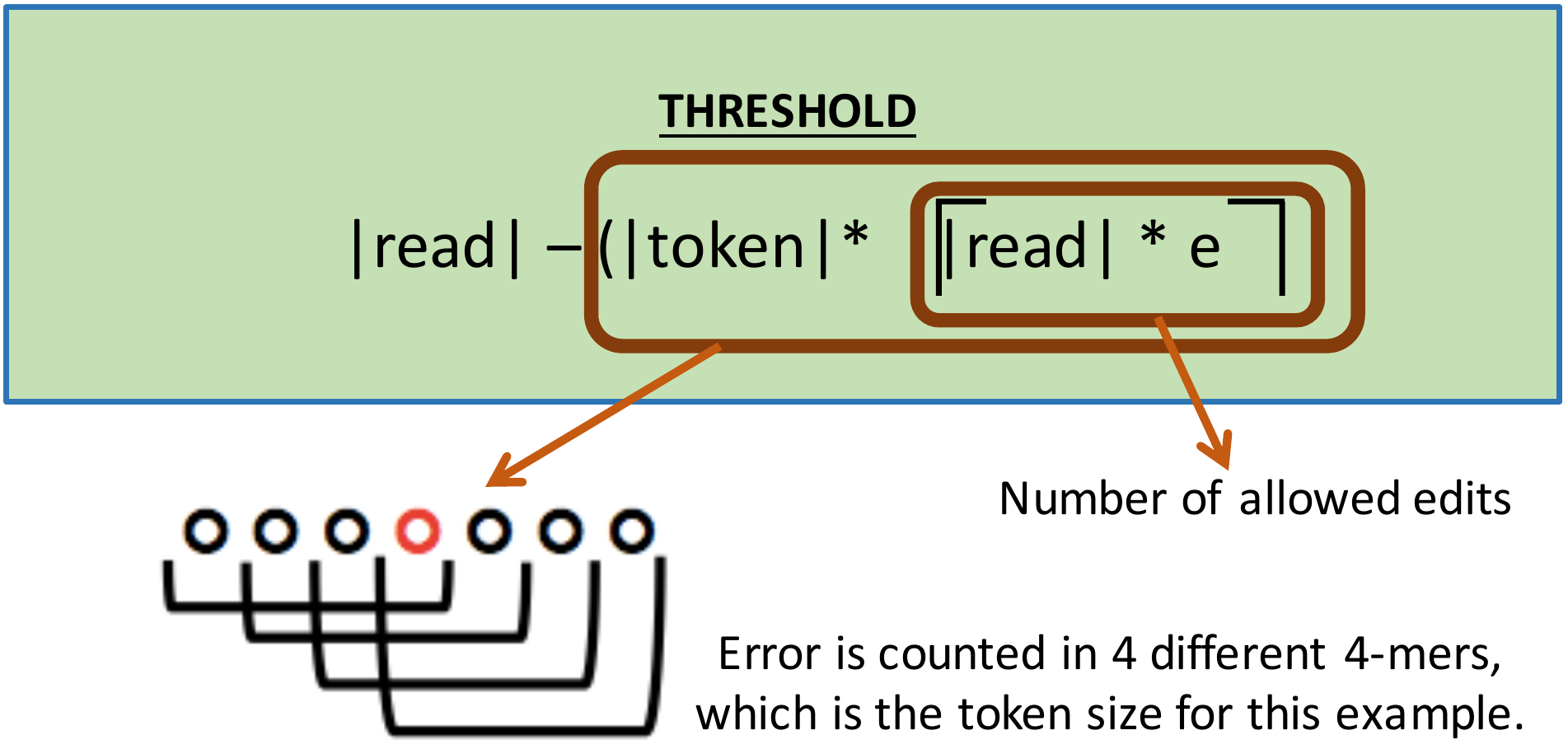}
  \caption{Threshold equation to find the threshold value of the summation of the tokens' existence values in the bit vectors. In this example, when $q=4$, every error propagates to the subsequent 3 tokens.}
  \label{fig:Threshold_eq}
\end{figure}

If reads mapped perfectly to the reference sequence, the threshold would simply
be the total number of tokens in a read or $read\_length~-~(n-1)$. However, due
to the need for allowing some differences in an alignment, we must compare the
accumulation sum against a lower value taking into account the worst case error
rate. This threshold can be calculated using the equation given in
Figure~\ref{fig:Threshold_eq}. As shown in the figure, a token of size $n$ in a
bin overlaps with $n$ other tokens. Assuming a single substitution error
between the read and reference sequence, the error will propagate to the $n$
previous tokens, meaning that those tokens may not be found in that bin. We
determine that the equation in Figure~\ref{fig:Threshold_eq} reflects the worst
case error distribution and error rate (e.g., an error rate of 5\% or less of
the read length is widely used~\cite{ahmadi2012hobbes, cheng2015bitmapper,
hatem2013benchmarking, xin2015shifted}) in a good match.  In the worst case,
where the maximum number of errors occurs and every error affects the $n$
adjacent tokens, the valid accumulation threshold is at its lowest value. 

After comparing the accumulated sum against the threshold calculated using the
appropriate values (read size, error rate threshold, and token size), GRIM-Filter
returns control to the read mapper to align those locations that pass the
filter. This process is repeated for all locations, which significantly reduces the
number of alignment operations and ultimately reducing the end-to-end read
mapping runtime. 

\subsection{Candidacy for 3D-stacked Memory Implementations}

We identify \textbf{three} characteristics of GRIM-Filter that make it a great
candidate implementing in for 3D-stacked memory: 1) only requires very simple
operations, 2) highly parallelizable since each bin can be operated on
independently, and 3) it is highly memory-bound requiring a single memory
access for approximately every three instructions.
 
\section{Mapping to 3D-Stacked Memory}

In this section, we first introduce \emph{3D-stacked DRAM} and describe how
GRIM-Filter can be easily mapped to utilize this emerging technology, which
attempts to bridge the disparity between processor speed and memory bandwidth.
As this disparity increases, memory becomes more of a bottleneck in the
computing stack~\cite{mutlu2003runahead}. Along with 3D-stacked DRAM, which
enables much higher bandwidth and lower latency compared to conventional DRAM,
the disparity between processor and memory is alleviated by the re-emergence of
\emph{Processing-in-Memory}, which integrates processing units inside or near
the memory system to leverage high in-DRAM bandwidth and reduce energy
consumption by reducing the amount of data transferred to the processor.  In
this section, we briefly explain the required background for these two
technologies, which we will leverage to accelerate DNA read mapping.


\subsection{3D-Stacked Memory}

\begin{figure}[h]
    \centering
    \includegraphics[width=0.6\linewidth]{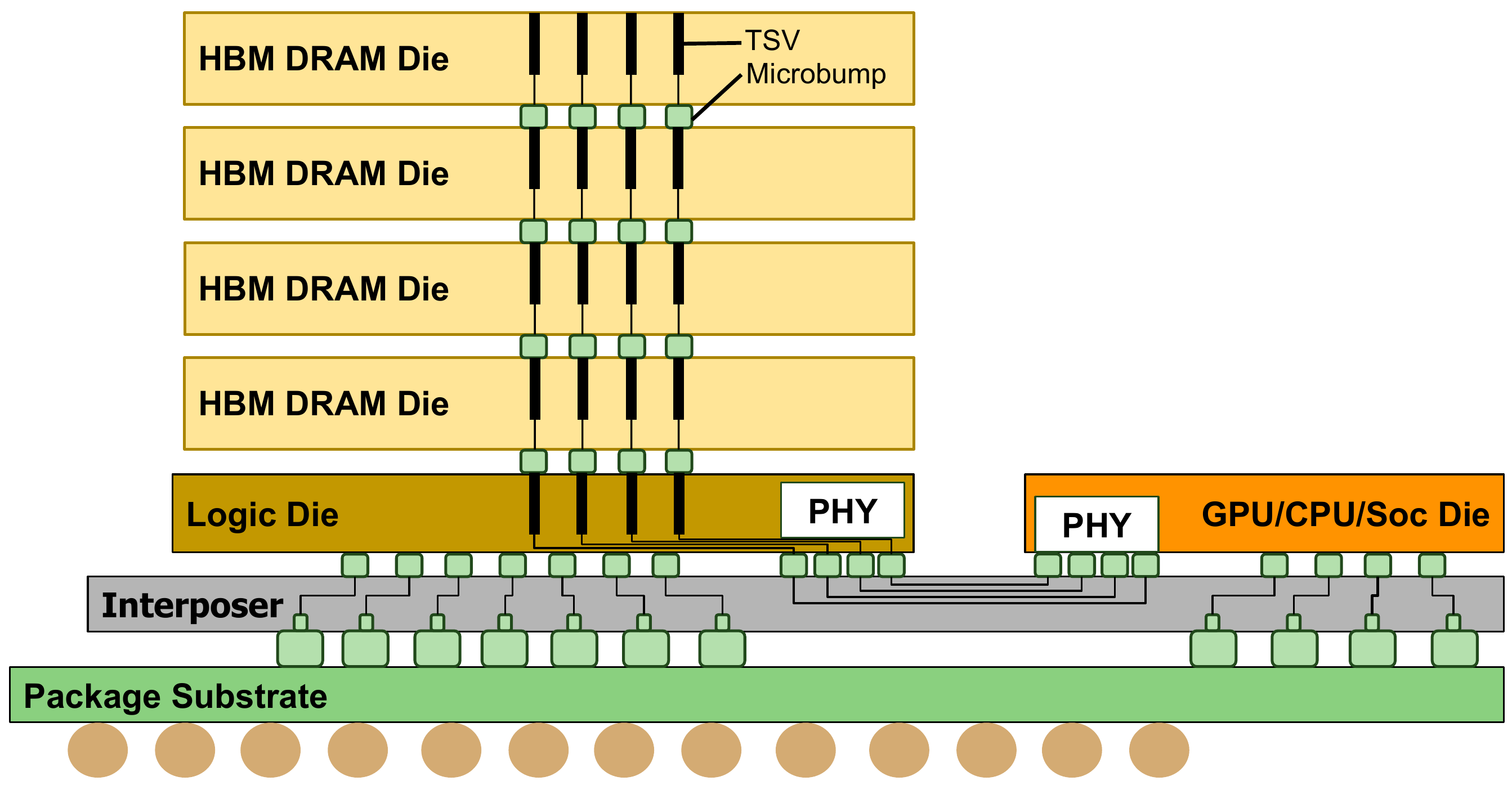}
    \caption{3D-Stacked DRAM example: High Bandwidth Memory consists of stacked memory layers and a logic layer connected by high bandwidth TSVs and microbumps~\cite{AMD-HBM}. The 3D-stacked memory is then connected to an SOC with an interposer layer that provides high bandwidth between the logic layer and the processing units on the package substrate.}
    \label{fig:HBM}
\end{figure}

3D-stacked DRAM has a much higher internal bandwidth than conventional DRAM,
thanks to the closer integration of logic and memory using
\emph{through-silicon via} (TSV) technology as seen in Figure~\ref{fig:HBM}.  TSVs
are vertical interconnects that can pass through the silicon wafers of a 3D
stack of dies \cite{kim2009study}.  TSVs have a much smaller feature size than
a standard interconnect, which enables a 3D-stacked DRAM to integrate hundreds
to thousands of these wired connections between stacked layers. Using these
wide wired connections, 3D-stacked DRAM can transfer bulk data simultaneously,
enabling much higher bandwidth compared to conventional DRAM.
Figure~\ref{fig:HBM} shows a 3D-stacked DRAM (High Bandwidth Memory, HBM
\cite{AMD-HBM}) based system that consists of a 4-layer stacked DRAM using
TSVs, a processor die, and silicon interposer that connects the stacked DRAM
and the processor. The vertical connections in the stacked DRAM are very wide
and very short which results in {\em high bandwidth} and {\em low power
consumption}, respectively~\cite{lee2015simultaneous}. There exist many
different 3D-stacked DRAM architectures available today. High Bandwidth Memory
(HBM) is already integrated into the new AMD Radeon$^{TM}$ R9 Series Graphics
Cards~\cite{AMD-R9-Graphics} and NVIDIA also announced that they will use HBM
in their future products~\cite{o2014highlights}.  Hybrid Memory Cube (HMC) is
also being developed by a number of different contributing companies
\cite{altera-HMC-UG, HMC-sources}.  Other new technologies are also around the
corner and can enable processing-in-memory, such as Micron's Automata Processor
(AP) \cite{dlugosch2014efficient} and Tibco transactional application servers
\cite{tibco-imc, micron-automata}.

{\bf Processing-in-Memory.} A key technique to increase the memory system
bandwidth and reduce energy consumption in the memory system is placing
computation units inside the memory system (e.g., PIM). Today, we see
processing capabilities appearing in or near conventional
DRAM~\cite{ahn2015scalable, ahn2015pim, lee2015simultaneous, seshadrifast,
seshadri2013rowclone, seshadri2015gather}. By enabling computation within or
near the memory system and only transferring the results to the CPU, PIM
provides significant performance improvements and energy reductions compared to
the conventional system architecture that transfers all data to the process and
only executes instructions within the CPU \cite{ahn2015scalable, ahn2015pim,
akin2015data, guo20143d}.

{\bf 3D-stacked DRAM with PIM.} Combining these two new technologies,
3D-stacked DRAM and PIM enable great opportunities to build very high
performance systems. A popular architecture for proposed 3D-stacked DRAM
consists of multiple stacked memory layers and a logic layer that control the
stacked memory, as shown in Figure~\ref{fig:HBM}. As many prior works
show~\cite{ahn2015scalable, ahn2015pim, akin2015data, guo20143d, loh20083d,
zhu20133d}, the logic layer in 3D-stacked DRAM can be utilized not only for
managing the stacked memory layers, but also for integrating
application-specific accelerators. Since the logic layer already exists and has
enough space to integrate compute units, integrating application-specific
accelerators in the logic layer requires very small design and implementation
overhead and little to no hardware overhead. 3D-stacked DRAM architecture
enables us to fully customize the logic layer for the acceleration of
applications~\cite{ahn2015pim, zhu20133d}.

\subsection{Mapping GRIM-Filter} 

We use mrFAST with FastHASH~\cite{xin2013accelerating} as our baseline for code
and performance. Our bit vector based implementation exists as an extension to
FastHASH as a simple series of calls to an Application Programming Interface
(API). FastHASH has an inflexible set of parameters, so there is not as much
system-specific tuning that can be done. However, for shared data structures
between FastHASH and GRIM-Filter, all parameters are kept consistent for a fair
comparison. For those parameters specific to the bit vectors data structure in
GRIM, we run tests to find a set of parameters that result in a highly
effective filter for our system (shown in Section~\ref{sec:results-analysis}). 

Due to the simplicity of our bit vector algorithm, we claim a low development
and area cost for the logic layer in the 3D-stacked memory device. The required
hardware for the logic layer as seen in Figure~6 simply 
depends upon the bandwidth available directly from the memory layers via TSVs.

\begin{figure}[H]
    \label{fig:3DDRAM-GRIM}
    \centering
    \includegraphics[width=\linewidth]{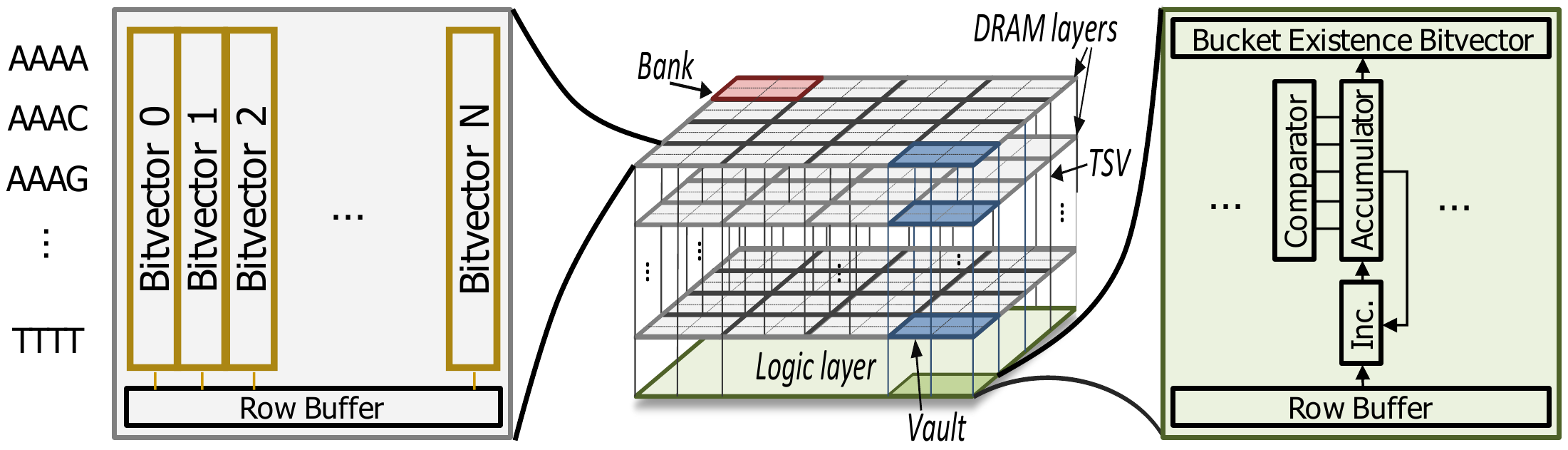}
    \caption{3D-stacked DRAM~\cite{akin2015data} with our proposed customized logic layer as well as proper data distribution throughout the memory layers. The logic layer is comprised of the row buffer, which represents the number of bits that can be transferred from the memory layers at once. Each bit in the row buffer has its own incrementer, accumulator, and comparator. The bit vectors are organized in memory so that multiple bins can be checked against the filter in parallel.}
\end{figure}

GRIM-Filter involves reading $p$ bits (\emph{existence bits}) in parallel from
differing bins representing the bin existence for the same token. We distribute
our bit vectors throughout memory such that {\em 1)} every bit representing the
existence of a given token across all bins is allocated a contiguous region of
memory, and {\em 2)} all bits describing a given bin $n$ from bit vectors of
different tokens will fall in the same column.  We then use these existence
bits to increment the accumulator in the corresponding indices and repeat for
all tokens in the read. This summation step simply requires a vector of
incrementers, where each sum value is represented by $\lceil\log_2
(read\_size)\rceil$ bits. The maximum value that the final sum can be is
equivalent to the size of the read simply due to the fact that that is the
number of tokens that compose each read. The number of required sum values and
incrementers is specified by $p$.  After this has been repeated for all tokens
in the read, we can reference the accumulators and compare the value to the
required threshold to determine whether to discard a location. 

In order to simplify referencing the accumulator, we utilize comparators for
each of the accumulators after summing across each token. We can reduce the
final accumulator values to a Boolean representing whether or not the read
could possibly exist in the bin. Depending on the available bandwidth of the
memory module in question, we simply require $p$ incrementer lookup tables
(LUT), $p$ 7-bit counters (for our particular sets of 100 base pair reads), $p$
comparators, and a single $(num\_bins)$-bit vector that holds the final result
for the given read at each bin. As future 3D-stacked memory devices are
expected to have more parallelism, the hardware overhead increases linearly,
but the performance overhead of GRIM-Filter reduces equally. GRIM-Filter
requires a very small and simple logic layer which gives it an edge over other
filtering algorithms that could be implemented on the logic layer.

\section{Experimental Methodology}

{\bf Evaluated Read Mapper.} We evaluate our proposal using the
state-of-the-art hash table based read mapper mrFAST with
FastHASH~\cite{xin2013accelerating}. We chose this mapper for our evaluations
as it provides high accuracy in the presence of large error rates, which is
required to detect genomic variants within and across
species~\cite{alkan2009personalized, xin2013accelerating}. However, we note
that GRIM-Filter can be used with any other mapper.

{\bf Major Evaluation Metrics.} We evaluate {\em 1)} the false negative rate
(i.e., proportion of locations that pass the filter, but result in a poor
match) of our GRIM-Filter, and {\em 2)} the performance improvement of the
end-to-end read mapper when using GRIM-Filter. To obtain both results, we first
integrate GRIM-Filter into mrFAST with FastHASH~\cite{xin2013accelerating}. We
measure the false negative rate of our filter (and the baseline filter used by
the mapper) as the ratio of the number of locations that passed the filter but
did not result in a mapping over all locations that passed the filter. We
detail how we measure the performance improvement of our mechanism next.

{\bf Performance Evaluation.} We measure the execution time improvement of our
mechanism by taking three measurements: {\em 1)} execution time of the baseline
mapper without GRIM-Filter (obtained by executing the source code of the
mapper, which is available as open source~\cite{xin2013accelerating}), {\em 2)}
execution time of the baseline mapper with GRIM-Filter's software
implementation, which does not take advantage of emerging memory technologies
(obtained by executing the source code of the baseline mapper integrated with
our software version of GRIM-Filter, which we will make available as open
source software), {\em 3)} execution time of the baseline mapper with
GRIM-Filter, which takes advantage of execution on 3D-stacked memory. To obtain
the last entity, we measure the execution time of the software GRIM-Filter
segments in mrFAST and subtract this from the obtained execution time in 2.
Then, using a validated in-house simulator similar to
Ramulator~\cite{kim2015ramulator}, we determine the overhead in offloading
GRIM-Filter to a 3D-stacked memory system and add the overhead execution time
to obtain the final execution time. We chose this methodology to estimate the
runtime of GRIM-Filter on 3D-stacked memory technologies as such technologies
that perform in-memory computation are unavailable to us at this point in time.

{\bf Evaluation System.} Our evaluation system is an Intel(R) Core$^{TM}$
i7-2600 CPU @ 3.40GHz with 16 GB of RAM for all experiments.

{\bf Data Sets.} We used ten real data sets from the 1000 Genome Project Phase
1 1000 Genomes Project Consortium (2012). These were the same data sets used by
Xin, et al~\cite{xin2013accelerating} for a fair comparison.
Table~1 lists the read length and size of each data set.

 \addtocounter{figure}{-1}
\begin{table*}[!htbp]\centering
\footnotesize 
{\begin{tabular}{@{}lcllllll@{}}\toprule
&& ERR240726\_1 & ERR240727\_1 & ERR240728\_1 & ERR240729\_1 & ERR240730\_1\\\hdashline
No. of Reads && 4031354 & 4082203 & 3894290 & 4013341 & 4082472 \\
Read Length && 100 & 100 & 100 & 100 & 100\\\toprule
&& ERR240726\_2 & ERR240727\_2 & ERR240728\_2 & ERR240729\_2 & ERR240730\_2\\\hdashline
No. of Reads && 4389429 & 4013341 & 4013341 & 4082472 & 4082472 \\
Read Length && 100 & 100 & 100 & 100 & 100\\\toprule
&&&&&&
\end{tabular}}{}
\label{Tab:benchmark} 
\centering\caption{Benchmark data, obtained from the 1000 Genomes Project Phase I~\cite{1000GP2012}}
\end{table*}


\section{Sensitivity Analysis and Results}
\label{sec:results-analysis} 

\begin{figure*}[h]
    \centering
    \includegraphics[width=0.9\linewidth]{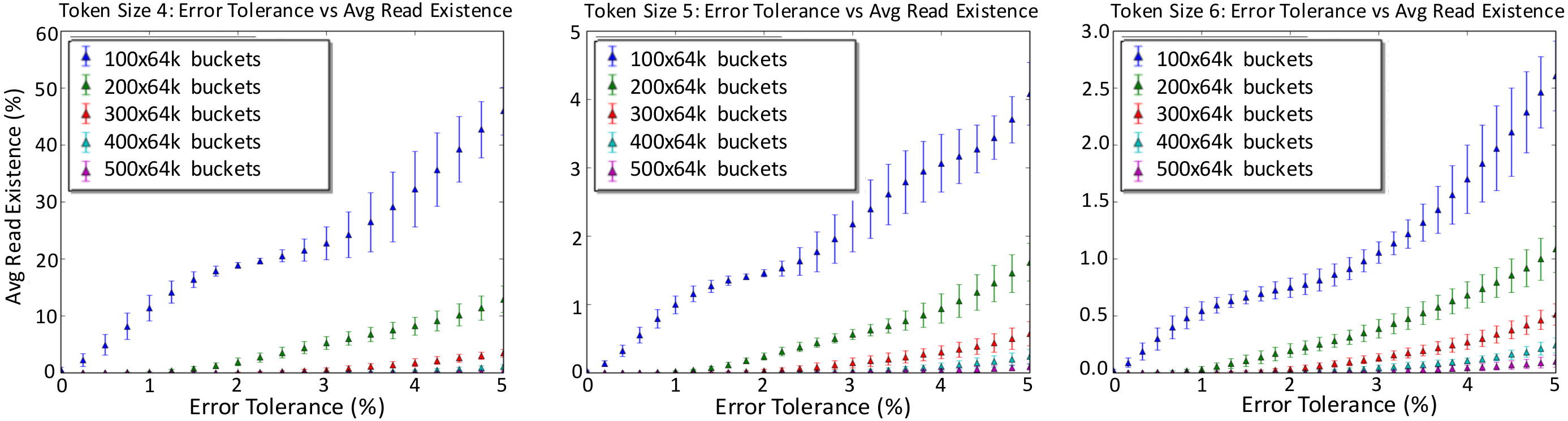}
    \caption{Token lengths 4-6 Read Existence: We use a representative set of reads to collect this data. We see how as error tolerance increases, the average read existence (the ratio of bins that must be checked with alignment to all bins comprising the genome) increases. We note that this value reflects the filter's effectiveness, assuming that the filter eliminates all false negatives (a low value is good, but zero would mean that the filter eliminates all candidate mappings). Our error tolerance reflects the threshold value that we use to determine whether we can filter any given bin. From Figure~\ref{fig:Threshold_eq}, we can see how varying the \% error tolerance would affect the threshold that we use for filtering. Because we take the $ceil$ of the read length multiplied with error tolerance (e), we see that partial error tolerances would change the threshold calculations even in the case of 100 bp reads. We can see how the plots marked with different colors show that as we increase the number of bins, the existence ratios decreased. The three plots represent the values found as we varied the token size.}
    \label{plot:substr4}
\end{figure*}

We first profiled the reference human genome in order to determine a range of
parameters that were reasonable to use for GRIM-Filter.  We were able to
determine the points of diminishing returns for several parameter values. This
data is presented in Section~\ref{subsec:motivational-data}. Using this
preliminary data, we could reduce the required experiments to a reasonable
range of parameters. Our implementation enabled the variation of runtime
parameters (number of bins, token size, error threshold, etc.) within
the ranges of values that we determined from our preliminary experimentation
for the best possible results.  We then were able to quantitatively evaluate
the improvements in the false negative rate and runtime over mrFAST with
FastHASH. Our results for the full mapper with GRIM-Filter are presented in
Section~\ref{subsec:data}.

\subsection{Parameter Evaluation Results}
\label{subsec:motivational-data} 

In order to determine a range for the parameters that we used for
experimentation, we ran a series of analyses on the fundamental characteristics
of the human reference genome. Our initial experiments were designed to
determine the \emph{memory footprint} of our algorithm for effective
performance improvements. To show how each of the different parameters affect
the performance of GRIM-Filter, we study a preliminary sweep on the parameters
with a range of values that would not incur excessive amounts of memory.
Figure~\ref{plot:substr4} shows how varying a number of different parameters
affects the \emph{average read existence} across the bins. We define average
read existence to be the ratio of bins that pass the filter to all bins
comprising the genome, for a representative set of reads. We want this value to
be as low as possible because it reflects the filter's ability to filter
incorrect mappings. The fewer bins that these reads, in the representative set,
map result in possible mappings, the more likely it will be that we will not
have to align a given location. Across the three plots, we vary the token size from
4 to 6. Within each plot, we vary the number of bins to split the reference
genome into, denoted by the different colors. The x-axes varies the error
threshold between a match, and the y-axes show average read existence. We plot
the average and min/max across our 10 data sets (Table~1) as
indicated respectively by the triangle and whiskers. We make \emph{three}
observations.  First, we observe, across the three plots, that increasing the
\emph{token size}, from 4 to 5, shows massive drops in the read existence while
5 to 6 exhibits significantly diminishing returns. This is due to the fact
that, given a random pool of A,C,T,G's, the probability of observing a
substring of size $q$ is $({\frac{1}{4}})^q$.  However, due to the non-uniform
distribution of base pairs across the genome and the bin sizes, we see
diminishing returns on the average read existence. Second, we observe that
across the plots, each increase in the number of bins results in a decrease in
the read existence.  This is understandable due to the fact that the bin size
decreases as the number of bins increases and for smaller bins, we have a
smaller sample size that any given substring could exist within. When sweeping
the number of bins, we use multiples of 64k because it is an even multiple of
the number of TSVs between the logic and memory layers in today's 3D-stacked
memories. We want to use a multiple of 64k so that we can utilize all TSVs for
each access. Third, we observe that for each plot, increasing the error
threshold results in an increase in the read existence.  This is due to the
fact that if we allow errors, a wider variety of sequences map to the same
read.  We conclude from this figure that using tokens of size 5 gives the best
tradeoff between memory consumption and filtering efficiency. 

\begin{figure}[h]
    \centering
    \includegraphics[width=0.4\linewidth]{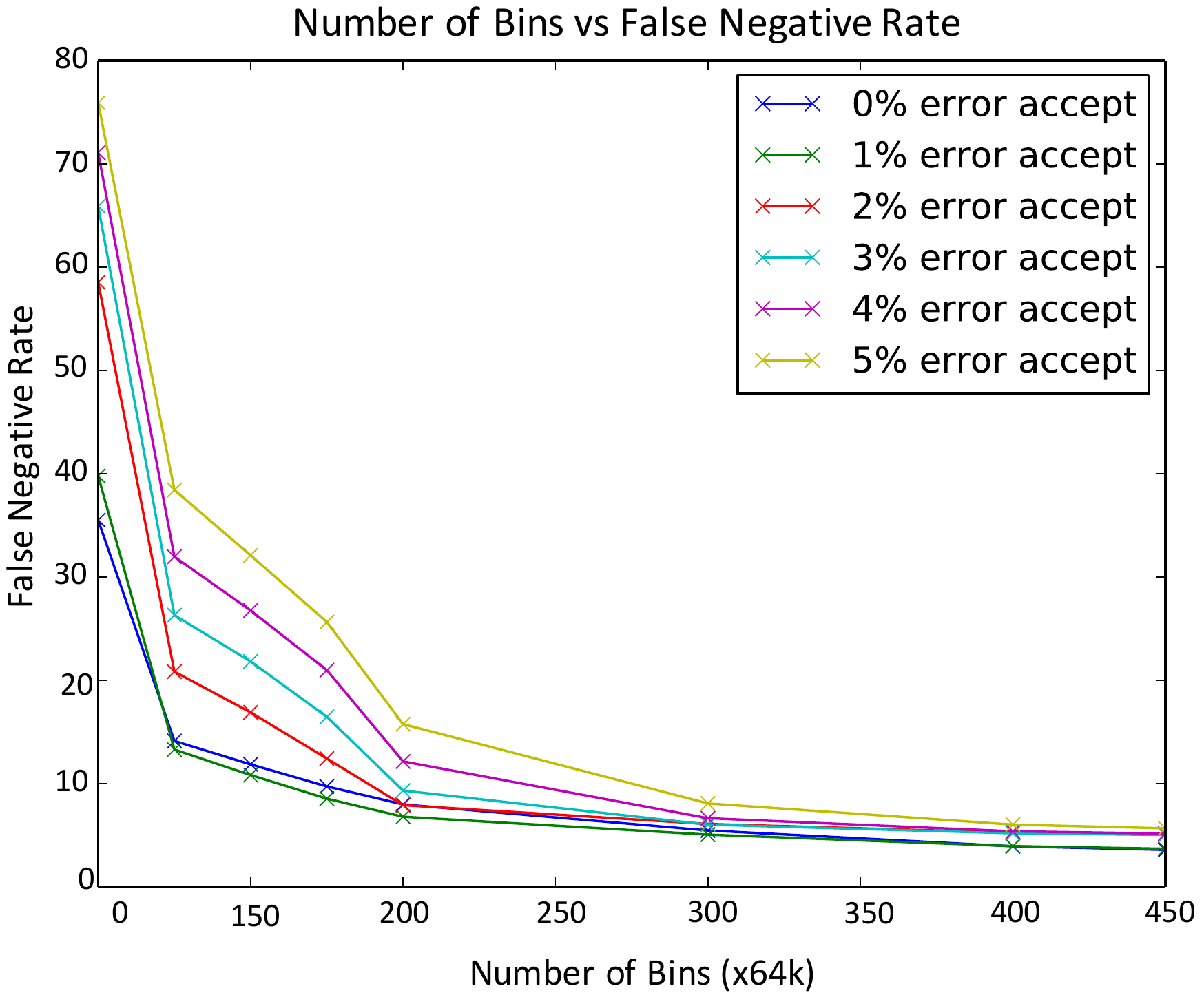}
    \caption{Bins Parameter Sweep: When running the actual experiments for a subset of the entire benchmarks, we find our filter's false negative rate as we vary different parameters. We see that as we increase the error threshold, we see a consistent increase in the false negative rates regardless of the number of bins used. We also note that increasing the number of bins results in diminishing returns around 300x64k bins over all error thresholds.}
    \label{plot:buckets_FPR}
\end{figure}

To show how we chose our final bit vector size to use for experimentation, we
sweep the number of bins and the error threshold (e\%).
Figure~\ref{plot:buckets_FPR} shows how varying these parameters affects the
false negative rates of the filtering algorithm. The x-axis varies the number
of bins, while the different colors represent the different error thresholds.
We make two observations from this plot. First, we find that with more bins
(thus a smaller bin size), the false negative rate (i.e., the number of
locations that pass through the filter, but don't result in a mapping after
alignment) decays exponentially. After 300x64k bins, we start to see
diminishing returns on the reduction of false negatives for all error
thresholds. Second, we observe that as we increase the error threshold, we see
that regardless of the other parameters, the false negative rates increase.
However, we can take advantage of the convergence of the different error
thresholds depending on the number of bins. After approximately 300x64k bins,
we see very small differences in the false negative rates for differing error
thresholds. Due to the slight improvements for false negative rates with an
increasing number of bins and the fact that number of bins minimally affects
the runtime of our filtering algorithm, we choose a value that reflects a
reasonable memory footprint given the other parameters.  We conclude that
employing 450x64k bins results in the best tradeoff between memory consumption,
filtering efficiency, and runtime. We note that the time to generate the bit
vectors is not included in our final runtime results because they only need to
be generated once per genome, either by the user or by the distributor.
However, for a better sense of the timescale, we find that with a genome length
$L$, we can generate the bit vectors in $9.03e-08 * L$ seconds when using
$450*64k$ bins (this is approximately 5 minutes for the human genome). 

Because the increasing number of bins results in more bit vectors, we must keep
this parameter at a reasonable value in order to retain a reasonable memory
footprint. Since we have chosen a token size of 5, we will require $t$ bit
vectors with a length of $4^5=1024$, where $t$ equals the number of bins we
segment the reference genome into. We choose $450x64k$ bins as a reasonable
tradeoff between memory footprint and false negative rate. This set of
parameters result in a total memory footprint of approximately $3.8$ GB for
storing the bit vectors of this mechanism, which is a reasonable size for
today's 3D-stacked memories. 

We ran several experiments to examine the benefits behind GRIM-Filter's ability
to parallelize consecutive bins. We noticed significant benefit in exploiting
parallelism when $p$ is 4096 (which is the bandwidth for HBM2)
\cite{o2014highlights}. In approximately 10\% of the $k$-mers, we see a
significant decrease (98.6\%) in required window retrievals. In the remaining
$k$-mers, we see approximately 10-20\% decrease in required window retrievals.
From HBM2 specifications \cite{o2014highlights}, we note that the available
bandwidth between memory and logic layer is 4096 bits, therefore our chosen
experimental $p$ value was 4096. Given larger $p$ values, we have experimental
data showing a continual reduction of required window retrievals.


\subsection{Full Mapping Results} 
\label{subsec:data} 

We used a popular seed-and-extend mapper, mrFAST~\cite{alkan2009personalized},
to retrieve all candidate mappings from ten real data sets from the 1000 Genome
Project Phase I~\cite{1000GP2012}. Table~1 lists the number of reads and size
of each read in each benchmark. In our experiments we use a token of length 5
and 450x64k bins as discussed in Section~\ref{subsec:motivational-data}.

\begin{figure}[h]
    \centering
    \includegraphics[width=0.7\linewidth]{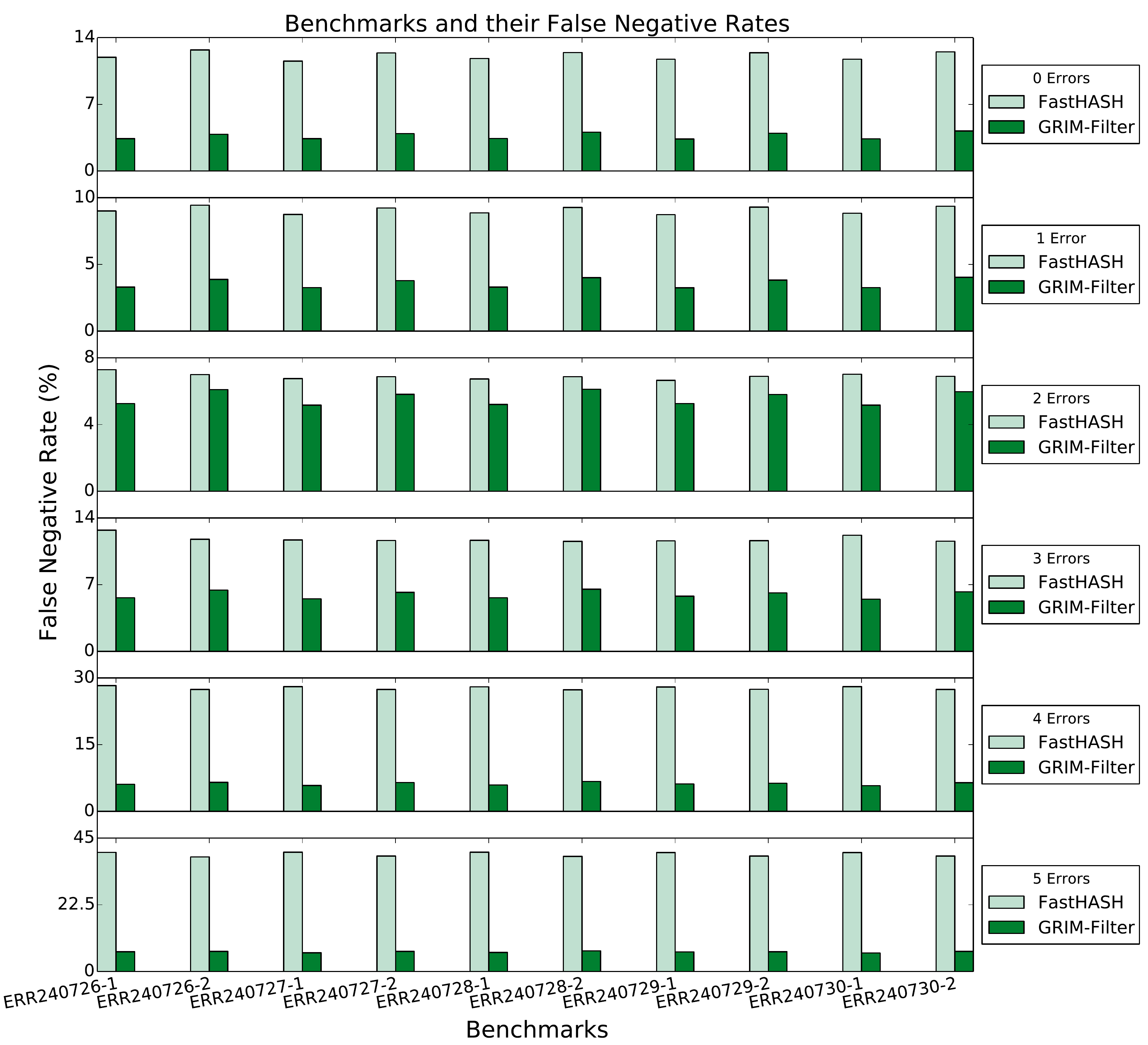}
    \caption{False Negative Rates for GRIM-Filter: comparison of GRIM-Filter vs. FastHASH (both used with mrFAST) abilities to discard locations that alignment would have resulted in a poor match, as error threshold varies.}
    \label{plot:FPR}
\end{figure}

Figure~\ref{plot:FPR} shows the number of false negative locations that pass
through GRIM-Filter compared to the baseline. The shared x-axis indicates the ten
sets of reads and the y-axis indicates the false negative rate. The light green
and dark green respectively mark the baseline and GRIM-Filter, and the graphs
descending vary in error thresholds. We make two observations. First,
we note a significantly lower false negative rate for all benchmarks in all
ranges of error thresholds when compared to the results of FastHASH.
Second, we observe a phenomenon where the false negative rates increase when
increasing the error thresholdfrom 0\% to 2\% and then decrease from
3\% to 5\%.  We attribute this to a combination of factors. This includes the
fact that increasing the error threshold results in more
acceptable mapping locations. However, the number of candidate locations do not
change. This naturally results in a smaller false negative rate. There is
another underlying factor: as acceptable error threshold increases, our
thresholding value decreases and allows for more locations to pass through the
filter resulting in an increased false negative rate. We conclude that the
interaction of these two factors are the cause for the initial increase and
later decrease in the false negative rates. We note that when using this filter
for higher error threshold, we observe larger improvements in
the false negative rate which can be reflected in the runtime.  When
comparing our filtering algorithm to FastHASH for an error threshold of
5\%\footnote{It is most important to compare to 5\% error threshold as it is
the accepted worst case error rate for read mappers and provides the highest
sensitivity.}, we see that our algorithm results in \emph{5.97x} less false
negative locations on average across the benchmarks. This is reflected directly
as a decrease in the end-to-end runtime, since fewer locations must be fully
aligned.
\begin{figure}[h]
    \centering
    \includegraphics[width=0.7\linewidth]{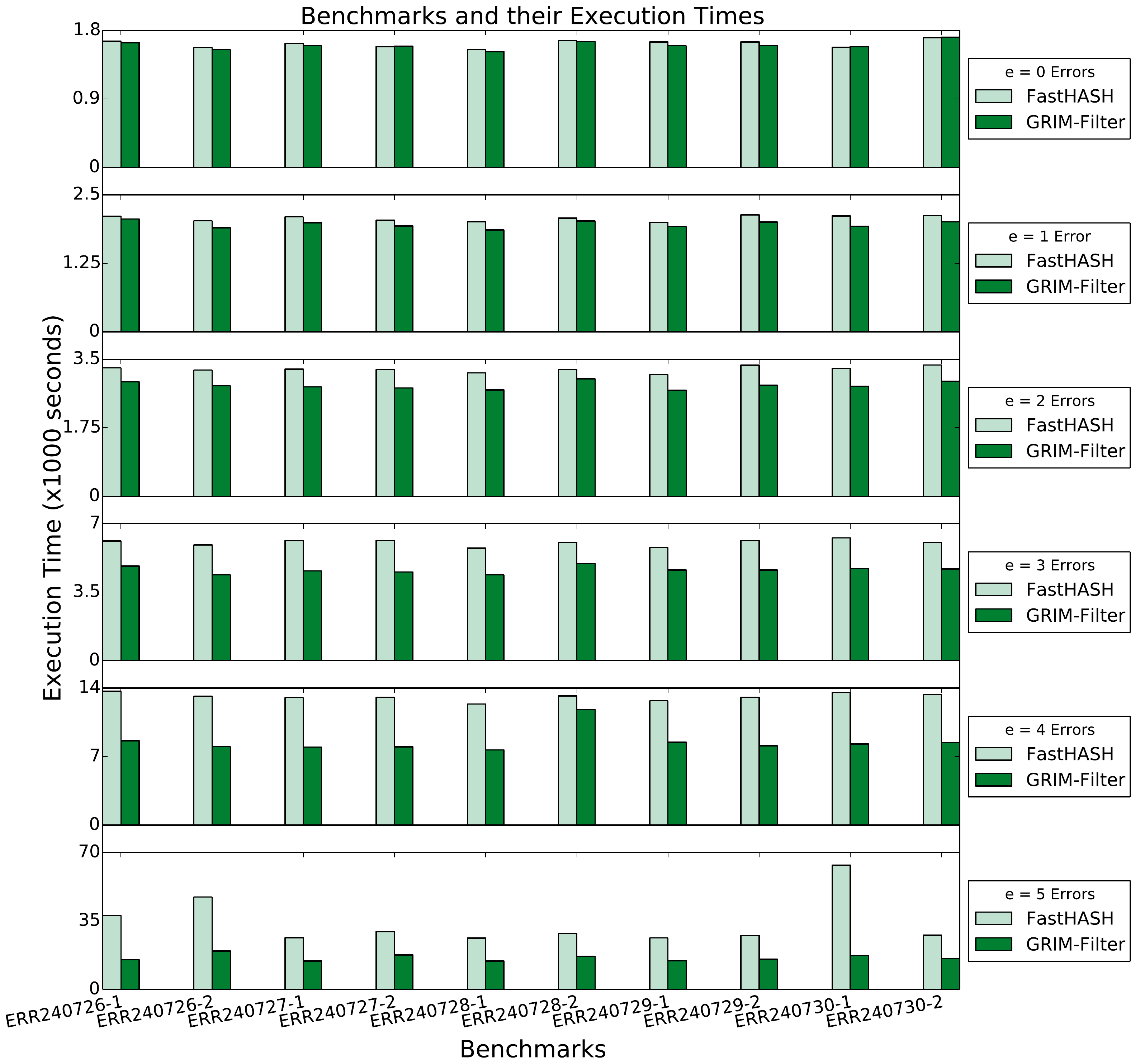}
    \caption{GRIM-Filter Runtimes: We compare end-to-end runtimes of mrFAST with GRIM-Filter against mrFAST with FastHASH. This figure shows the comparison of their runtimes across the benchmarks as error threshold varies.}
    \label{plot:runtime}
\end{figure}

Figure~\ref{plot:runtime} compares the execution time of GRIM-Filter against
the baseline, mrFAST with FastHASH. This graph follows the same format as the
previous, except the y-axis now represents the execution time scaled to 1000
seconds. We make two observations. First, we observe that GRIM-Filter shows
performance improvement over all benchmarks regardless of the error threshold.
Second, as the error thresholdincreases, we gain increasingly more performance
benefit.  This is due to the fact that GRIM-Filter is able to discard many more
locations than FastHASH at higher error thresholds, thus saving much more
execution time by ignoring unnecessary alignments. Again, because of the
importance of high sensitivity for calling structural variations and the direct
correlation between runtime and error threshold, we report all numbers only
looking at the maximum error threshold of 5\%. When we compare the runtime of
mrFAST with GRIM-Filter against the previous fastest read mapper, mrFAST with
FastHASH, we find a 2.08x (3.65x) performance boost on average (max) across the
benchmarks. When we further break down the computation time, we find that our
performance gains are from an average decrease across datasets of 83.7\%
computation time on false negative locations. We conclude that employing
GRIM-Filter can significantly enhance the performance of a state-of-the-art
mapper. 

\section{Related Works} 
\label{sec:related} 

To our knowledge, this is the first paper to exploit 3D-stacked DRAM and its
processing-in-memory capabilities to overcome the recent bottleneck shift to
memory bandwidth in read mapping due to the immense improvement on the prior
bottleneck, alignment. In this section, we briefly describe related works that
aim to accelerate read mapping with hardware support. 

Many prior works used FPGAs to accelerate alignment. These
include~\cite{aluru2014review, arram2013hardware, arram2013reconfigurable,
ashley2010clinical, chiang2006hardware, hasan2007hardware, houtgast2015fpga, 
mcmahon2008accelerating, olson2012hardware, papadopoulos2013fpga,
waidyasooriya2014fpga} and all accelerate read mapping using
customized FPGA implementations of different existing read mapping algorithms.
Arram et al.~\cite{arram2013reconfigurable} accelerate SOAP3 tool on an FPGA
engine and it shows up to \textit{134x} speedup compared to BWA.  Houtgast et
al.~\cite{houtgast2015fpga} present a FPGA-accelerated version of BWA-MEM that
is \textit{3x} faster compared ot its software implementation.  Other works use
GPUs~\cite{blom2011exact, liu2012soap3, luo2013soap3, manavski2008cuda} for the
same purpose. Liu et al.~\cite{liu2012soap3} accelerate BWA and Bowtie by
\textit{7.5x} and \textit{20x}, respectively. However, all these accelerators
are still bottlenecked by memory bandwidth. Compared to these accelerators, our
approach overcomes the memory bandwidth bottleneck by utilizing the
up-and-coming 3D-stacked DRAM with a newly designed algorithm that is specific
to this technology.

In the case of other hardware optimized implementations with much higher
speedups, they focus on the acceleration of the actual alignment (the dynamic
programming step) which is the bulk of the computation in mapping. While many
works have managed to attain the maximum possible acceleration in alignment
through multiple iterations of implementations ranging from ASIC to FPGA, we
explore a newer area in mapping that requires significantly less computation.
We show that we can accelerate the entire mapping pipeline by utilizing the
inherent massive parallelism in 3D-DRAM. We note that GRIM-Filter is orthogonal to
other filters and mapper steps and can be stacked on top of other existing
optimizations for further potential acceleration. We show that when we run
mapping with GRIM-Filter on a commodity CPU, we see $1.81x-3.65x$ performance
improvement. We speculate substantial potential in tying together the
implementation of this filter with other hardware optimized aligners.


\section{Conclusion} 

We introduced a new algorithm, GRIM-Filter, for accelerating genome read
mapping. GRIM-Filter takes advantage of an emerging technology, 3D-stacked
memory, which enables the efficient use of processing-in-memory to overcome the
memory bottleneck in read mapping today. We utilize the processing-in-memory
capability of 3D-stacked technology and exploit its massive internal bandwidth
to run GRIM-Filter which efficiently and quickly filters large segments of the
genome for later steps of read mapping. With the most relevant alignment error
acceptance rate of 5\%, we show that GRIM-Filter filters locations with
approximately \emph{5.59x--6.41x} smaller false negative rate than FastHASH and
performs \emph{1.81x--3.65x} faster than the fastest read mapper, mrFAST with
FastHASH. GRIM-Filter is a universal filter that can be applied to any read
mapper.

We believe there is huge potential in adapting DNA read mapping algorithms to
state-of-the-art and emerging memory and processing technologies. With our
results, we hope that our paper, which introduces the first work in doing so
for 3D-stacked memories, which are increasingly common in today's computing
landscape, provides inspiration for other such works to design new sequence
analysis algorithms that take advantage of 3D-stacked memory.

\begin{backmatter}

%


\bibliographystyle{bmc-mathphys} 
\bibliography{references}      

\end{backmatter}
\end{document}